\documentclass[aps,prd,nofootinbib]{revtex4-2}

\usepackage{graphicx}
\usepackage{latexsym}
\usepackage{amsmath}
\usepackage{amsfonts}
\usepackage{amssymb}
\usepackage{verbatim}
\usepackage[utf8]{inputenc}
\usepackage[T1]{fontenc}
\usepackage{appendix}

\usepackage[colorlinks=true, pdfstartview=FitV, linkcolor=blue, citecolor=red, urlcolor=magenta]{hyperref}

\newcommand{\be}{\begin{equation}}
\newcommand{\ee}{\end{equation}}
\newcommand{\bea}{\begin{eqnarray}}
\newcommand{\eea}{\end{eqnarray}}

\newcommand{\ben}{\begin{eqnarray}}
\newcommand{\een}{\end{eqnarray}}

\begin{document}

\title{Casimir Effect for a Massive Scalar Field in Lorentz-Violating Aether Compactification}


\author{$^{1}$K. E. L. de Farias}
\email{klecio.lima@uaf.ufcg.edu.br}

\author{$^{1,5}$M. A. Anacleto}
\email{anacleto@df.ufcg.edu.br}

\author{$^{1,2,4}$ A. A. Ara\'{u}jo Filho}
\email{dilto@fisica.ufc.br}

\author{$^{3}$M. B. Cruz}
\email{messiasdebritocruz@servidor.uepb.edu.br}

\author{$^{1}$R. A. Dantas}
\email{robsonalvesdantas@gmail.com}

\author{$^{1}$Amilcar R. Queiroz}
\email{amilcarq@df.ufcg.edu.br}

\author{$^{1,5}$E. Passos}
\email{passos@df.ufcg.edu.br}

\affiliation{$^{1}$Departamento de F\'{\i}sica, Universidade Federal de Campina Grande,\\
Caixa Postal 10071, 58429-900, Campina Grande, Para\'{\i}ba, Brazil.}
\affiliation{$^{2}$Departamento de F\' isica, Universidade Federal da Para\' iba,\\  Caixa Postal 5008, Jo\~ ao Pessoa, Para\' iba, Brazil.}
\affiliation{$^{3}$Centro de Ciências Exatas e Sociais Aplicadas (CCEA), Universidade Estadual da Paraíba (UEPB), R. Alfredo Lustosa Cabral, s/n, Salgadinho, Patos - PB, 58706-550 - Brazil.}
\affiliation{$^{4}$Center for Theoretical Physics, Khazar University, 41 Mehseti Street, Baku, AZ-1096, Azerbaijan.}
\affiliation{$^{5}$Unidade Acad\^emica de Matem\'atica, Universidade Federal de Campina Grande,\\ 58429-970,  Campina Grande, Para\'{\i}ba, Brazil.}

\begin{abstract}
This work investigates the influence of Lorentz symmetry breaking, introduced by an aether-like field $\alpha_\phi$, on the Casimir effect within a five-dimensional flat spacetime. By considering a quasiperiodic condition regulated by the parameter $\beta$ and an extra dimension compactified at scale $b$, we derive closed-form expressions for the Casimir energy and the resulting force between two parallel plates under Neumann boundary conditions. Our results demonstrate that $\beta$ acts as a crucial control parameter, enabling a continuous transition between attractive and repulsive regimes, with a characteristic symmetry around $\beta = 0.5$. We show that the Lorentz-violating parameter $\alpha_\phi$ functions as an enhancement factor, significantly amplifying the vacuum interaction, while the geometric ratio $a/b$ proves decisive for system stabilization. Specifically, we find that the high-compactification regime leads to a plateau in the Casimir force, effectively stabilizing the interaction. Furthermore, we analyze the mass spectrum of the field, recovering standard geometric forms in the massless limit and demonstrating that while light fields ($M \ll 1$) exhibit subtle quadratic corrections, heavy fields ($M \gg 1$) lead to an exponential suppression of the Casimir effect. The interplay between Lorentz violation and extra-dimensional compactification provides a rich mechanism with potential applications in the modulation of vacuum-induced interactions at micro and nano scales.
\end{abstract}


\maketitle


\section{Introduction} \label{intro}

The investigation of possible extra dimensions in the modern physics model has acquired notoriety as a solution for the open problems present in the Standard Model \cite{Arkani-Hamed:1998jmv}. These challenges include quantum gravity — as attempts to quantize the gravitational field imply the existence of additional dimensions as a consistency condition \cite{Arkani-Hamed:1998sfv} — as well as the nature of dark matter \cite{Dvali:1999tq, Bedroya:2025fwh, Carvalho:2026osd} and dark energy \cite{Greene:2007xu, Ferrer:2005hr} that permeates the Universe. In this sense, Theodor Kaluza \cite{Kaluza:1921tu} and Oskar Klein \cite{Klein:1926tv} proposed a theory (KK theory) that postulates the existence of a compactified fifth dimension in an attempt to unify electromagnetism and gravity.

The imposition of boundary conditions, whether by physical plates or compact geometry, leads to the phenomenon theorized by H. B. Casimir in 1948 \cite{Casimir:1948dh}. Originally, this effect described the vacuum state of the electromagnetic field confined between two large, parallel, conducting plates. These boundaries restrict vacuum excitations to specific wavelengths, generating a force that was experimentally confirmed with high precision in the following decades \cite{Sparnaay:1958wg, Lamoreaux:1996wh, Lambrecht:2000zz}. In the context of extra dimensions, this concept extends to the topological Casimir effect \cite{Petrosyan:2025xwg}, where non-trivial topology and periodicity conditions along compact dimensions modify the fluctuation spectrum of quantum fields, analogous to the effect of physical plates. Consequently, this phenomenon results in shifts in the vacuum expectation values of physical observables that depend directly on the local geometry and the compactification scheme, serving as an essential mechanism for the stabilization of moduli fields and for the understanding of dark energy \cite{Milton:2002hx, Elizalde:2006iu, Wongjun:2013jna}.

Despite the theoretical relevance of extra dimensions, a significant challenge arises regarding the compactification scale $b$. In conventional KK theory, $b$ is predicted to be near the Planck scale ($10^{-35}$ m), rendering its detection nearly impossible with current experimental energies \cite{Bailin:1987jd}. To circumvent this limitation, Carroll and Tam proposed "aether compactification" as a mechanism to allow for larger extra dimensions without necessitating brane-world scenarios to mediate the five-dimensional bulk's influence on the observable Universe \cite{Carroll:2008pk}. This approach, which inherently involves the violation of Lorentz symmetry (LV), not only assists in the stabilization of moduli fields but is also investigated as a potential origin for the dark energy responsible for the cosmic accelerated expansion \cite{Milton:2002hx}.

Unlike braneworld models, this compactification scheme does not generate corrections to the four-dimensional Newton's inverse-square law \cite{Carroll:2008pk}, ensuring a consistent effective theory. This approach is particularly relevant given the challenges of unifying General Relativity with Quantum Mechanics — a central goal of contemporary physics - pursued by frameworks such as String Theory and Loop Quantum Gravity \cite{Zwiebach:2004tj, Rovelli:2004tv}. In these scenarios, LV emerges as a potential consequence, either through the introduction of tensor fields that define a preferential direction in spacetime, as in Standard-Model Extension models \cite{Kostelecky:1988zi, Kostelecky:2003fs}, or through space-time anisotropy, as proposed in Ho\v{r}ava-Lifshitz gravity \cite{Horava:2009uw}. Consequently, aether fields — by LV within the extra dimensions — alter the mass splitting of the Kaluza-Klein towers, modifying the spacing between quantum states. Although such violations are expected to occur predominantly at the Planck scale, residual signatures of this anisotropy may be observable in low-energy phenomena. This could, in principle, render these dimensions accessible to a four-dimensional observer, making the study of the Casimir effect an ideal laboratory for investigating such quantum manifestations.

Several studies on the aether field have been conducted recently \cite{Carroll:2008pk, Obousy:2008xi, Obousy:2008ca, Chatrabhuti:2009ew, Furtado:2016zqi, Santos:2017yov}. In this regard, one of the most frequently employed and simplest models involves an extra dimension compactified into a circle of length $b$ \cite{Greene:2007xu, Feng:2013zza, Erdas:2021fwe}. The existence of this additional dimension allows for indirect detection through phenomenological manifestations in the vacuum energy density, which can be modified to become appreciable at a physically significant scale. As previously discussed, the Casimir effect provides a method to probe this energy, where the extra dimension's influence may either enhance or suppress the force's magnitude depending on the system's quantum fluctuations. This configuration leaves a signature in the vacuum energy between the parallel plates, offering insights into the existence of extra dimensions and their underlying nature as investigated in Refs. \cite{deFarias:2023xjf,Rohim:2025ial}, which they use experimental results to find phenomenological constraints on the extra dimension values associated with the system. Various other investigations within this theme have explored the impact of LV on the Casimir effect \cite{Cruz:2017kfo, Cruz:2018bqt, Cruz:2018thz, Cruz:2020zkc, deMello:2022tuv}. Furthermore, the stabilization of extra dimensions considering LV parameters—such as the aether field—has been analyzed in \cite{Obousy:2008ca, Chatrabhuti:2009ew}.

Within this line of investigation, our work explores the Casimir effect in a Kaluza-Klein scenario where the scalar field is subject to distinct boundary conditions. Specifically, we examine the influence of parallel plates under Neumann conditions, coupled with the compactification of the extra dimension through quasi-periodic constraints. This approach allows for a generalization of the vacuum energy behaviour across a continuous range of periodicities, spanning the spectrum between periodic and anti-periodic cases. Furthermore, it unveils the correlation between topology and the aether field, illustrating how their interaction modifies quantum fluctuations in this multidimensional manifold.

The present work is organized as follows: Section \ref{sec2} introduces the theoretical model, where the aether field is incorporated into the Kaluza-Klein framework, establishing the Neumann boundary conditions and quasi-periodicity constraints. In Section \ref{sec3}, we derive the Casimir energy and explore the impact of the extra-dimensional compactification and Lorentz violation. Section \ref{sec4} is dedicated to obtaining the Casimir force and analyzing its phenomenology. Finally, Section \ref{sec5} presents our concluding remarks. Throughout this article, we adopt natural units $\hbar=c=1$ and the metric signature $(-,+,+,+,+)$.

\section{Aether-scalar coupling}
\label{sec2}

\begin{figure}[!htb]
    \includegraphics[scale=0.35]{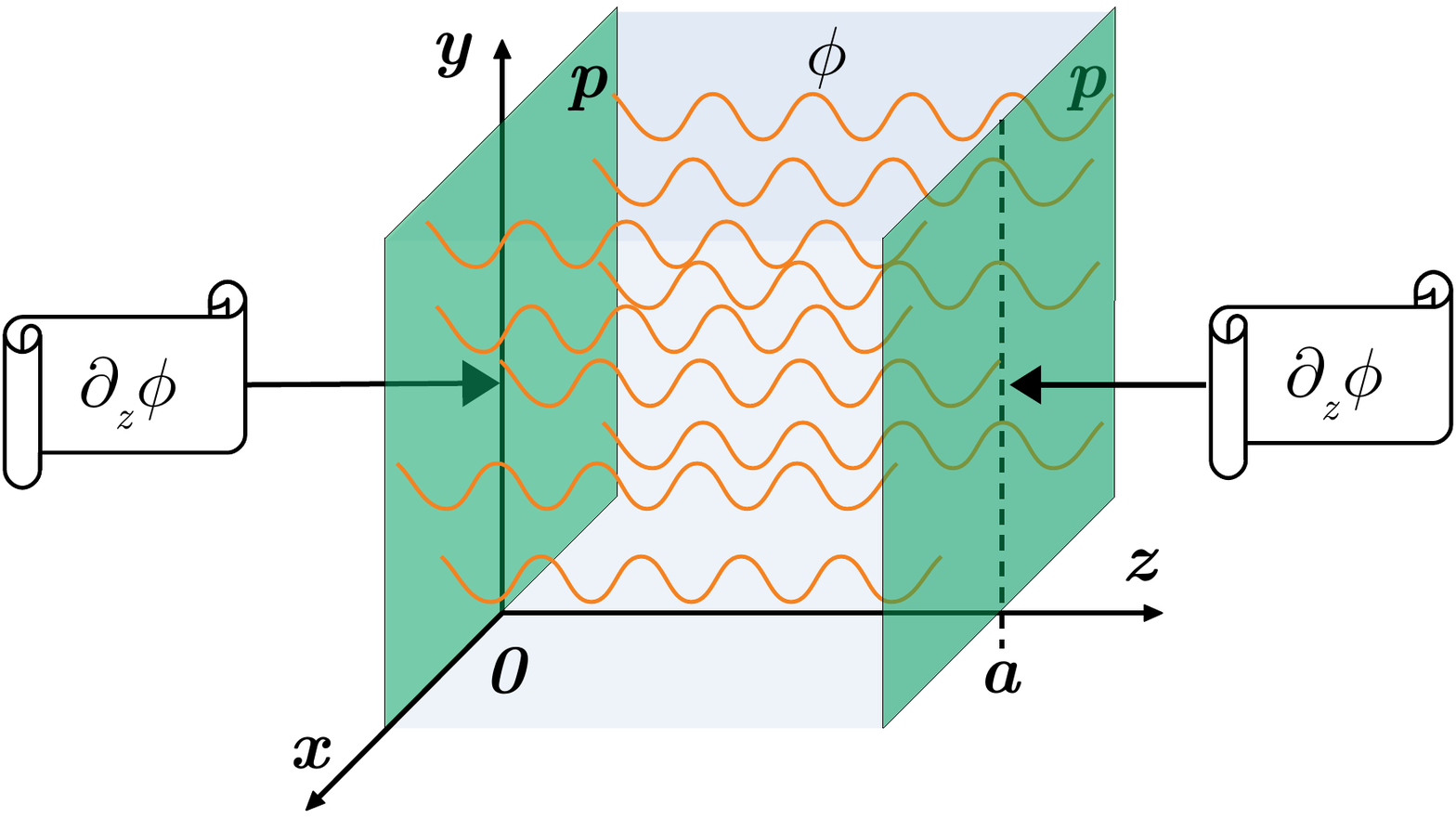}
    \caption{Schematic representation of the system geometry, featuring two parallel plates separated by a distance $a$ along the $z$-axis. Both plates are subject to Neumann boundary conditions.}
    \label{fig002}
\end{figure}

\begin{figure}[!htb]
    \includegraphics[scale=0.35]{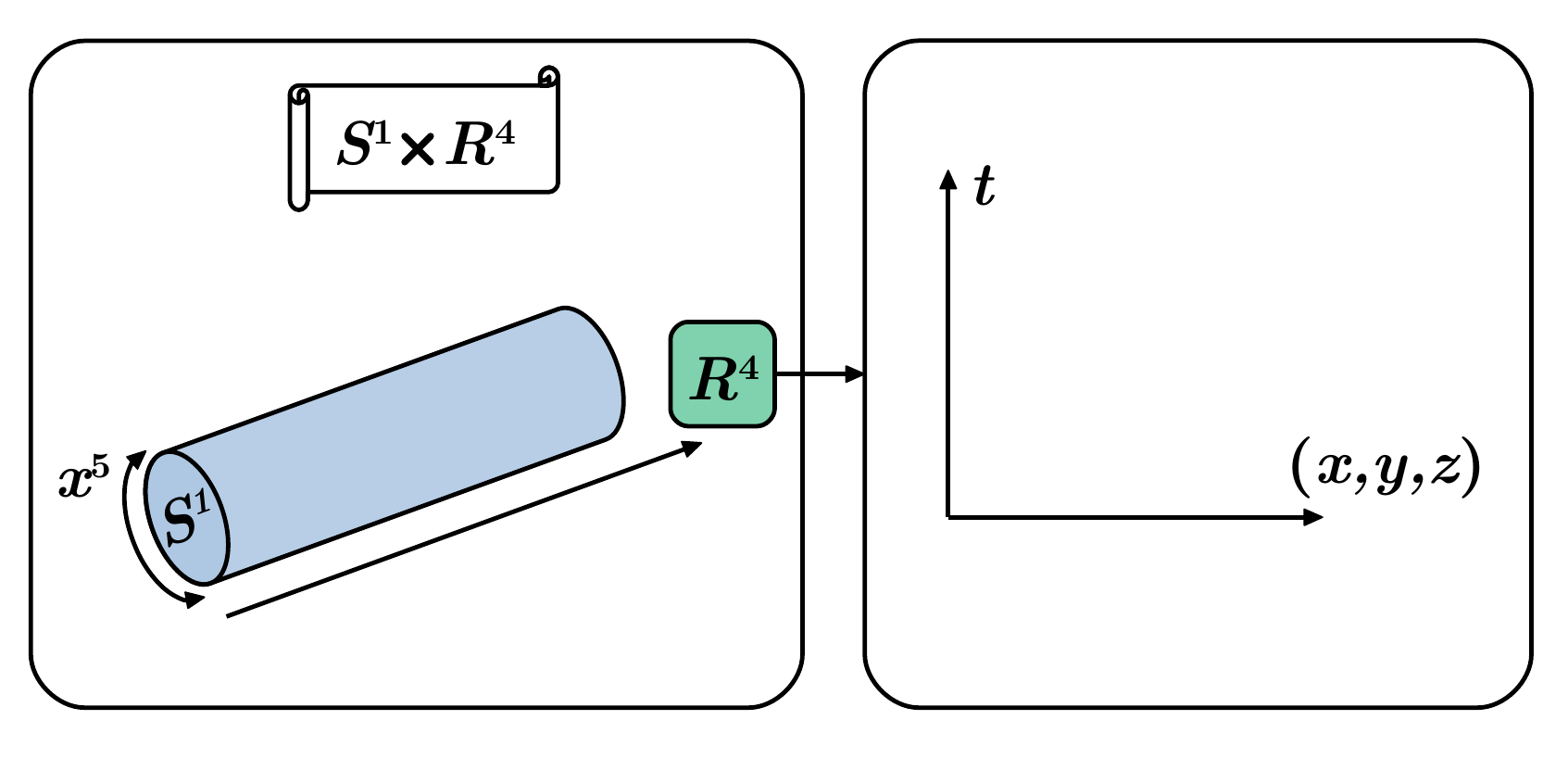}
    \caption{Schematic representation of the extra dimension $x^5$, compactified into a circle of scale $b$.}
    \label{fig001}
\end{figure}

In this section, we introduce the theoretical model in which the Klein-Gordon equation is modified within the context of the aether approach \cite{Colladay:1996iz, Colladay:1998fq}. Our goal is to derive a dispersion relation compatible with the Neumann boundary conditions applied to the fields on two parallel plates in the $z$-direction, as illustrated in Fig. \ref{fig002}. Here, the aether field is introduced by a five-dimensional spacelike vector $u^a=(0,0,0,0,v)$, where the extra-dimension coordinate, given by $x_5$, is compactified into a circle of radius $b$ (see the scheme in Fig. \ref{fig001}). This vector has mass dimension and is assumed to be constant. Hence, $u^a$ establishes a preferential direction, thereby breaking the Lorentz symmetry along the extra dimension.

In this context, the interaction between a real scalar field $\phi$ and the aether field, via a minimal coupling term, is described by the following Lagrangian:
\begin{equation}
    \mathcal{L}_\phi=-\frac{1}{2}\left(\partial_a \phi\right)^2-\frac{1}{2 \mu_\phi^2} u^a u^b \partial_a \phi \partial_b \phi - \frac{1}{2}m^2\phi^2,
    \label{mod_lagra}
\end{equation}
where $\mu_\phi$ is a coupling parameter of mass dimension. From Eq. \eqref{mod_lagra}, we obtain the modified Klein-Gordon equation, given by
\begin{equation}
    \partial_a \partial^a\phi + m^2\phi = -\mu_\phi^{-2} \partial_a\left(u^a u^b \partial_b \phi\right).
    \label{mod_klein-gordon}
\end{equation}
By considering the five-dimensional plane wave solution in the form $e^{i k_a x^a} = e^{i (k_\mu x^\mu + k_5 x^5)}$, and taking $u_a u^a = v^2$, the modified dispersion relation is given by:
\begin{equation}
    -k^\mu k_\mu = m^2 + \left(1 + \alpha_\phi^2 \right) k_5^2,
    \label{eq03}
\end{equation}
where $\alpha_\phi=v / \mu_\phi$ is a dimensionless parameter.

In this model, the compactification of the extra dimension is implemented by imposing a quasi-periodic boundary condition, which is described by
\begin{equation}
\phi(t, x, y, z, x^5)=e^{-2 \pi \beta i} \phi(t, x, y, z, x^5+b),
    \label{eq05}
\end{equation}
as well as the Neumann boundary conditions on the parallel plates, given by:
\begin{equation}
\partial_z\phi(t, x, y, z, x^5)_{z=0}=0,\left.\quad \partial_z \phi(t, x, y, z, x^5)\right|_{z=a}=0.
    \label{eq06}
\end{equation}
Applying these boundary conditions to the plane wave solution leads to the discretization of momenta in the $z$ and $x_5$ directions. Furthermore, the quasi-periodic boundary condition is characterized by a phase shift governed by the parameter $\beta$, within the interval $0 \leq \beta < 1$. Notably, this range encompasses both periodic ($\beta=0$) and anti-periodic ($\beta=1/2$) boundary conditions. Consequently, the extra dimension is compactified into a circle $\mathbb{S}^1$ of length $b$ via this quasi-periodic constraint.

The imposition of Neumann boundary conditions along the $z$-direction is particularly relevant, as it allows the model to be probed via the Casimir effect, which can be measurable in laboratory experiments. Consequently, the existence of an extra dimension may manifest as detectable shifts in the Casimir effect. In this framework, we consider an $\mathbb{R}^4 \times \mathbb{S}^1$ spacetime, where $\mathbb{R}^4$ denotes the standard Minkowski manifold and $\mathbb{S}^1$ represents the compactified fifth dimension. By applying both boundary conditions to this geometry, the associated dispersion relation is given by:
\begin{equation}
    -k^\mu k_\mu=k_j^2 + m^2 + \left(1+\alpha_\phi^2\right)k_n^2 \ \ \ \text{where} \ \ \ k_j=\frac{j \pi}{a},
        \label{eq07}
\end{equation}
where $j=0,1,2,\dots$ is associated with the Neumann boundary conditions, with $a$ being the distance between the plates, and 
\begin{equation}
    k_n=\frac{2 \pi}{b}(n+\beta) \ \ \ \text{with} \ \ \ n=0, \pm 1, \pm 2, \dots
\end{equation} 
is related to the quasi-periodic boundary condition, where $b$ is the compactification length. We should point out that the momentum discretization $k_j$ in Eq. \eqref{eq07} has the same form as the Dirichlet boundary condition, although in the latter case $j=1, 2,\dots$ . This difference stems from the field modes — sine for Dirichlet and cosine for Neumann — nonetheless, both yield the same physical energy \cite{ambjorn1983properties}. Consequently, the eigenfrequencies are written as:
\begin{equation}
    \omega_n^2=k_x^2+k_y^2+k_j^2+ m^2+ \left(1+\alpha_\phi^2\right)k_n^2.
    \label{EF1}
\end{equation}
The Casimir effect is evaluated in the following sections, starting from the dispersion relation provided in Eq. \eqref{EF1}.

\section{Casimir energy}
\label{sec3}

As previously mentioned, in flat spacetime, the Casimir energy arises as a natural consequence of the boundary conditions imposed on the quantum field by the presence of the plates. In the previous section, we established two distinct boundary conditions, as given in Eqs. \eqref{eq05} and \eqref{eq06}. Consequently, vacuum energy is expressed by the following expression:
\begin{equation}
  E_0=\frac{1}{2}\left(\frac{L}{2 \pi}\right)^{\ell} \int \mathrm{d}^{\ell} k \sum_j \sum_n \sqrt{k^2+k_j^2+ m ^2 +\left(1+\alpha_\phi^2\right)k_n^2},
  \label{e1}
\end{equation}
where $k^2 = k_x^2 + k_y^2$, $L^2$ is the area of the plates, and $\ell=2$ represents the dimension regularizer. We employ this parameter instead of the fixed dimension to evaluate the integral of the Eq. \eqref{e1} using the following identity:
\begin{equation}
    \int f(k) \mathrm{d}^n k=\frac{2 \pi^{n / 2}}{\Gamma(n / 2)} \int k^{n-1} f(k) \,\mathrm{d} k.
    \label{int_rela}
\end{equation}

Therefore, substituting Eq. \eqref{int_rela} into Eq. \eqref{e1}, we obtain the following expression for the vacuum energy:
\begin{align}
E_0&=\frac{1}{2}\left(\frac{L}{2 \pi}\right)^{\ell} \frac{2 \pi^{\ell / 2}}{\Gamma(\ell / 2)} \sum_j \sum_n \int \mathrm{d} k\, k^{\ell-1} [k^2+k_j^2+m^2+\left(1+\alpha_\phi^2\right)k_n^2]^{-r}\nonumber\\
& =\left(\frac{L}{2 \pi}\right)^{\ell} \frac{\pi^{{\ell} / 2} \Gamma(r-\ell/2)}{2 \Gamma(r)} \sum_j \sum_{n}\left\{\left(\frac{j \pi}{a}\right)^2+m^2 + \left(1+\alpha_\phi^2\right)\left[\frac{2 \pi}{b}(n+\beta)\right]^2\right\}^{\frac{\ell}{2}-r}.
\label{e1.1}
\end{align}
The parameter $r$ above is a regularization term that allows us to compute the integral above; the physical result is recovered in the limit $r \to -1/2$. By defining $s = r - \ell/2$, we obtain the following expression:
\begin{equation}
    E_0 =\left(\frac{2}{L}\right)^{2s+1} \frac{\pi^{s+1/ 2} \Gamma(s)}{2 \Gamma(-1/2)}\left(\frac{\pi}{a}\right)^{-2 s} \sum_{j=0}^{\infty} \sum_{n=-\infty}^{\infty}\left[\frac{a^2}{\tilde{b}^2}(n+\beta)^2+M^2 + j^2\right]^{-s},
\label{e2}
\end{equation}
where
\begin{equation}
\tilde{b}^2=\frac{b^2}{4\left(1+\alpha_\phi^2\right)} \ \ \ \ \ \text{and} \ \ \ \ \ \quad M=\frac{m a}{\pi}.
\label{LR}
\end{equation}
As argued in Ref. \cite{Carroll:2008pk}, the aether parameter $\alpha_\phi$ can reach values as large as $10^{15}$. In this sense, it acts as an effective length reduction according to Eq. \eqref{LR}. Now, we proceed to evaluate the summations in Eq. \eqref{e2}. First, we focus on the sum over $n$, to which we apply the Abel-Plana formula \cite{Saharian:2007ph}:
\begin{equation}
\sum_{n=0}^{\infty} f(n)=\int_0^{\infty} \mathrm{d} x f(x)+\frac{1}{2} f(0)+i \int_0^{\infty} \mathrm{d} x \frac{f(x)-f(-i x)}{e^{2 \pi x}-1}.
\label{e3}
\end{equation}
To apply the Abel-Plana formula, we first introduce the substitution:
\begin{equation}
    f(j)=\left[\frac{a^2}{\tilde{b}^2}(n+\beta)^2+M^2+j^2\right]^{-s}=\left[N^2+M^2+j^2\right]^{-s},
    \label{e4}
\end{equation}
where $N=a(n+\beta)/\tilde{b}$. The sum over $j$ in Eq. \eqref{e2} is given by
\begin{align}
\sum_{j=0}^{\infty} f(j) =\int_0^{\infty} \mathrm{d} x\left(N^2+M^2+x^2\right)^{-s}+\frac{(N^2+M^2)^{- s}}{2}+2 i^{1-2 s} \int_{\sqrt{N^2+M^2}}^{\infty} \mathrm{d} x \frac{\left(x^2-N^2 -M^2\right)^{-s}}{e^{2 x \pi}-1}.
\label{e5}
\end{align}
By applying the geometric series expansion:
\begin{equation}
    \frac{1}{e^{2x}-1} = \sum_{n=1}^{\infty} e^{-2nx},
    \label{eq:geometric_series}
\end{equation}
the second integral in Eq. \eqref{e5} yields:
\begin{equation}
    \int_{\sqrt{N^2+M^2}}^{\infty} \mathrm{d} x \frac{\left(x^2-N^2-M^2\right)^{-s}}{e^{2 x \pi}-1}=\sum_{q=1}^{\infty}\left(\frac{q}{\sqrt{N^2+M^2}}\right)^{-\frac{1}{2}+s} \pi^{-1+s} K_{-\frac{1}{2}+s}  \left(2 q \pi\sqrt{N^2+M^2} \right) \Gamma(1-s),
    \label{e6}
\end{equation}
where $K_{\alpha}(x)$ denotes the modified Bessel function of the second kind. So, by substituting Eq. \eqref{e6} into Eq. \eqref{e5}, the vacuum energy is expressed as:
\begin{align}
E_0= & \left(\frac{2}{L}\right)^{2s+1} \frac{\pi^{s+1/ 2} \Gamma(s)}{2 \Gamma(-1/2)}\left(\frac{\pi}{a}\right)^{-2 s} \sum_{n=-\infty}^{\infty}\left\{\int_0^{\infty} \mathrm{d} x\left(N^2+M^2+x^2\right)^{-s}+\frac{(N^2+M^2)^{- s}}{2}+\right.\nonumber \\
& \left.+2 i^{1-2s} \pi^{-1+s} \Gamma(1-s) \sum_{q=1}^{\infty}\left(\frac{q}{\sqrt{N^2+M^2}}\right)^{-\frac{1}{2}+s} K_{-\frac{1}{2}+s}(2 q \pi\sqrt{N^2+M^2})\right\} .
\label{e7}
\end{align}

Next, we evaluate the summation over $n$ in Eq. \eqref{e7}. Starting from the first term on the r.h.s., we evaluate the summation over $n$ by applying the identity in Appendix \ref{app:sum_identity} from Ref. \cite{deFarias:2023xjf}, which leads to:
\begin{align}
&\left(\frac{a}{\tilde{b}}\right)^{-2s}\!\!\!\! \int_0^{\infty}\!\!\!\! \mathrm{d} x\!\!\!\! \sum_{n=-\infty}^{\infty}\left[(n+\beta)^2+\frac{\tilde{b}^2}{a^2} (M^2+x^2)\right]^{-s}\!\!\!\!
 =\left(\frac{a}{\tilde{b}}\right)^{-2 s} \int_0^{\infty} \mathrm{d} x\left[2\tilde{b}^{-2 s}\int_0^{\infty}d\mu \left(\frac{\mu^2}{\tilde{b}^2}+\frac{M^2+x^2}{a^2}\right)^{-s}\right.  \nonumber \\
& \left.+\sin(\pi s)
\frac{4\Gamma(1-s)}{\pi^{1-s}}\left(\frac{\tilde{b}}{a} \sqrt{M^2+x^2}\right)^{\frac{1}{2}-s} \sum_{n=1}^{\infty} \frac{\cos (2 \pi n \beta)}{n^{\frac{1}{2}-s}} K_{\frac{1}{2}-s}\left(2 \pi n\frac{\tilde{b}}{a} \sqrt{M^2+x^2}\right)\right].
\label{e8}
\end{align}
Note that the integral over $\mu$ yields a divergent result and is typically renormalized \cite{Bordag:2009zz}. The integral over $x$, on the other hand, can be evaluated using the following identity \cite{Gradshteyn:1943cpj}:
\begin{equation}
\int_0^{\infty}\left(x^2+M^2\right)^{\nu / 2} K_\nu\left(c \sqrt{x^2+M^2}\right) \mathrm{d} x=\sqrt{\frac{\pi}{2}} \frac{1}{\sqrt{c}M^{\nu-\frac{1}{2}}} K_{\nu-\frac{1}{2}}(c M).
\end{equation} 
Thus, by applying Euler's reflection formula,
\begin{equation}
\Gamma(1-z) \Gamma(z)=\frac{\pi}{\sin \pi z} ,
\end{equation}
the Eq. \eqref{e8} can be rewritten as:
\begin{equation}
\left(\frac{a}{\tilde{b}}\right)^{-2s} \int_0^{\infty} \mathrm{d} x \sum_{n=-\infty}^{\infty}\left[(n+\beta)^2+\frac{\tilde{b}^2}{a^2} (M^2+x^2)\right]^{-s}=\left(\frac{a}{\tilde{b}}\right)^{-s} \frac{2\pi^{s}}{\Gamma(s)}  \sum_{n=1}^{\infty} \frac{\cos (2 \pi n \beta)}{M^{s-1}n^{1-s}}K_{s-1}(2\pi nM\frac{\tilde{b}}{a}).
\label{e11}
\end{equation}

An analytical expression for the second summation in Eq. \eqref{e7} can be derived by employing the recurrence relation \cite{Feng:2013zza}:
\begin{align}
\sum_{n=-\infty}^{\infty}\left(\frac{a}{\tilde{b}}\right)^{-2 s}(n+\beta)^{-2 s} =\left(\frac{a}{\tilde{b}}\right)^{-2s} \pi^{2s-1/2} \frac{\Gamma\left(\frac{1}{2}-s\right)}{\Gamma\left(s\right)} 2 \sum_{n=1}^{\infty}\frac{\cos (2 \pi n \beta)}{ n^{1-2s}} ,
\label{e12}
\end{align}
leading to
\begin{align}
    \sum_{n=-\infty}^{\infty} & \frac{\left(N^2+M^2\right)^{-s}}{2}=\frac{1}{2}\left\{\sqrt{\pi} \frac{\tilde{b}}{a} M^{1-2 s} \frac{\Gamma(-1 / 2+s)}{\Gamma(s)}+\right. \nonumber \\
    & \left.\left.+\sin (\pi s) \frac{4 \Gamma(1-s)}{\pi^{1-s}} \sum_{\ell=1}^{\infty}\left(\frac{\ell}{M}\right)^{-\frac{1}{2}+s}\left(\frac{a}{\tilde{b}}\right)^{-1 / 2-s} \cos (2 \pi \ell \beta) K_{\frac{1}{2}-s}\left(2 \pi \ell M \frac{\tilde{b}}{a}\right)\right]\right\}.
\end{align}
Therefore, substituting Eqs. \eqref{e11} and \eqref{e12} into Eq. \eqref{e7}, we arrive at the regularized vacuum energy:
\begin{align}
E_0 =&\left(\frac{2}{L}\right)^{2 s+1} \frac{\pi^{1 / 2}}{2 \Gamma(-1 / 2)} \frac{1}{a^{-2 s}}\left\{\left(\frac{a}{\tilde{b}}\right)^{-s} 2 \sum_{n=1}^{\infty} \frac{\cos (2 \pi n \beta)}{M^{s-1} n^{1-s}} K_{s-1}\left(2 \pi n M \frac{\tilde{b}}{a}\right)\right.\nonumber \\
& +2 \sum_{\ell=0}^{\infty}\left(\frac{\ell}{M}\right)^{-1 / 2+s}\left(\frac{a}{\tilde{b}}\right)^{-\frac{1}{2}-s} \cos (2 \pi \ell\beta) K_{1/2-s}\left(2 \pi \ell M \frac{\tilde{b}}{a}\right)\nonumber \\
& \left.+2 \sum_{q=1}^{\infty} \sum_{n=-\infty}^{\infty}\left(\frac{q}{\sqrt{N^2+M^2}}\right)^{-\frac{1}{2}+s} K_{-\frac{1}{2}+s}\left(2 q \pi \sqrt{N^2+M^2}\right)\right\}.
\end{align}
We now perform the renormalization by taking the limit $s \to -3/2$ and using the propriety $K_{-\alpha}(x)=K_{\alpha}(x)$. Thus, we obtain the following Casimir energy:
\begin{align}
E_C= & - \frac{L^2}{8a^3 }\left\{\left(\frac{a}{\tilde{b}}\right)^{3 / 2} \sum_{n=1}^{\infty} \frac{\cos (2 \pi n \beta)}{M^{-5 / 2} n^{5 / 2}} K_{\frac{5}{2}}\left(2 \pi n M \frac{\tilde{b}}{a}\right) + \frac{a}{\tilde{b}}\sum_{\ell=1}^\infty\left(\frac{\ell}{M}\right)^{-2} \cos (2 \pi \ell \beta) K_2\left(2 \pi \ell M \frac{\tilde{b}}{a}\right)\right.\nonumber \\
& \left.+ \sum_{n=-\infty}^{\infty} \sum_{q=1}^{\infty}\left(\frac{q}{\sqrt{N^2+M^2}}\right)^{-2} K_{2}\left(2 q \pi \sqrt{N^2+M^2}\right)\right\}\, ,
\label{casimir_ener_exac}
\end{align}
Note that, due to the quasiperiodic parameter $\beta$, the $n=0$ term in the last sum does not require regularization; this specific contribution will be further discussed in the limit $\beta \to 0$. The influence of the quasiperiodicity is manifest in each term within the brackets. Notably, the second term is independent of the plate separation $a$, yet it remains non-vanishing as $\beta \to 0$. This indicates that such a contribution arises solely from the compactified extra dimension, where the effects of Lorentz symmetry breaking are encapsulated in the redefined compactification scale $\tilde{b}$, which depends on the aether-like parameter $\alpha_\phi$.

\begin{figure}[h!]
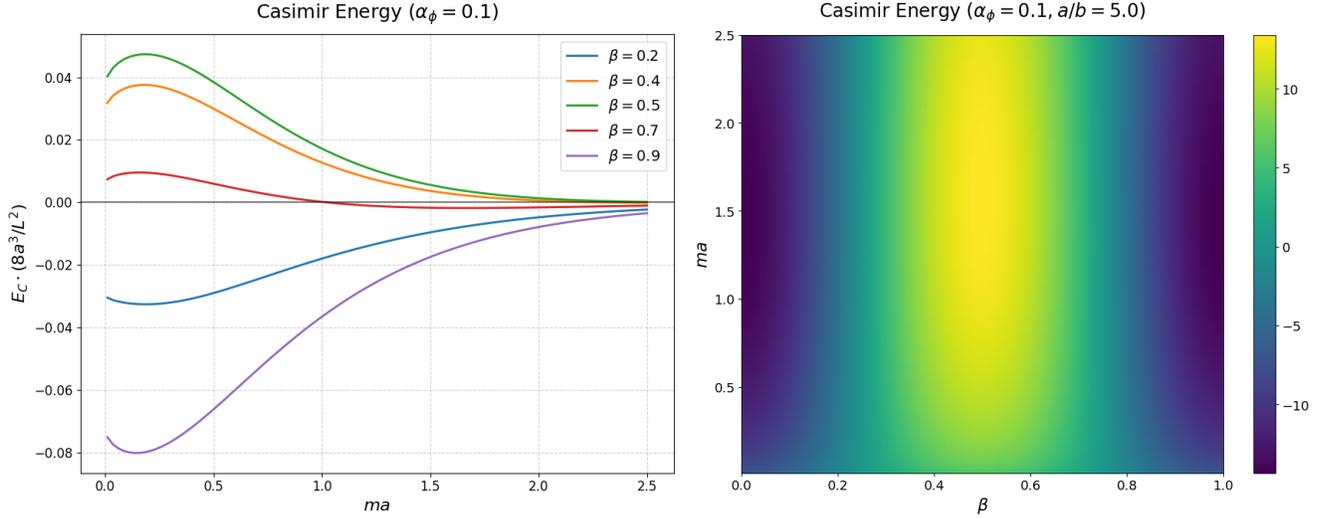

    \includegraphics[scale=0.40]{casimir_energy_beta.pdf}
    \includegraphics[width=0.465\linewidth]{casimir_energy_beta_hm.pdf}
    \caption{Casimir energy as a function of field mass $ma$ for different values of the periodicity parameter $\beta$. The energy exhibits a symmetric behavior around $\beta = 0.5$, with transitions between attractive and repulsive regimes occurring near $\beta = 0.2$ and $\beta = 0.7$.}
    \label{casimir_energy_beta}
\end{figure}

\begin{figure}[h!]
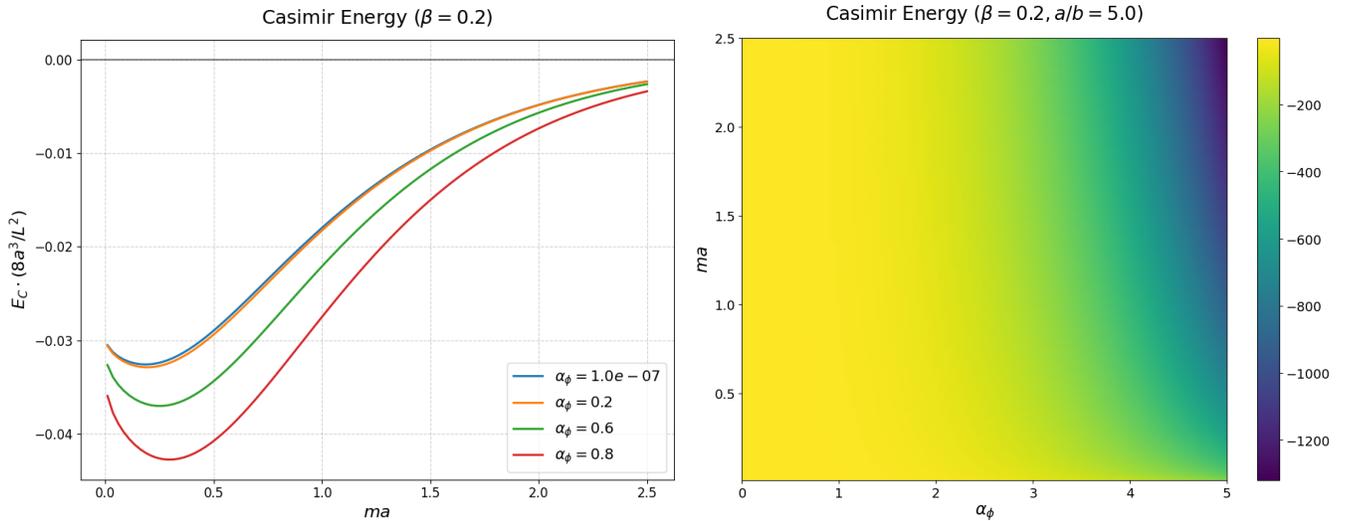

    \includegraphics[scale=0.40]{casimir_energy_alpha.pdf}
    \includegraphics[width=0.48\linewidth]{casimir_energy_alpha_hm.pdf}
    \caption{Casimir energy as a function of the mass parameter $ma$ for different values of the LV parameter $\alpha_\phi$, with $\beta = 0.2$. The plots show that increasing $\alpha_\phi$ enhances the magnitude of the vacuum energy, shifting the curves downward across the analyzed spectrum.}
    \label{casimir_energy_alpha}
\end{figure}

\begin{figure}[h!]
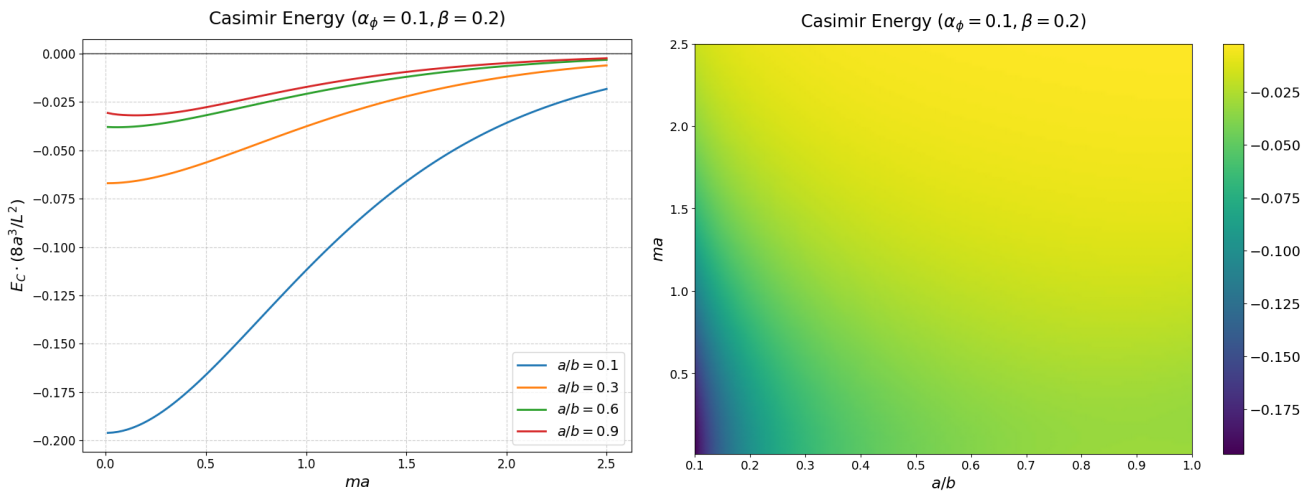

    \includegraphics[scale=0.375]{casimir_energy_ab.pdf}
    \includegraphics[width=0.49\linewidth]{casimir_energy_ab_hm.pdf}
    \caption{Casimir energy as a function for different geometric ratios $a/b$. The plot shows a sharp divergence in the limit of large extra dimensions ($a/b \to 0$) and an asymptotic stabilization of the energy density as the compactification scale becomes comparable to the plate distance.}
    \label{casimir_energy_ab}
\end{figure}

For a better visualization, we analyze the behavior of the dimensionless Casimir energy density, defined by Eq. \eqref{casimir_ener_exac}, as a function of the field mass ($ma$) under different parametric regimes. The influence of the periodicity parameter $\beta$ is presented in Fig. \ref{casimir_energy_beta}. It is observed that varying $\beta$ allows for the modulation of not only the intensity but also the sign of the Casimir energy, enabling attractive, repulsive, or even null configurations. A clear symmetry is observed around $\beta = 0.5$, which corresponds to the maximum positive (repulsive) energy value. Conversely, the values $\beta = 0$ and $\beta = 1$ result in the minimum negative (attractive) values. Interestingly, numerical analysis reveals that the energy becomes positive in the approximate interval of $0.2 < \beta < 0.7$, regardless of the $a/b$ ratio. In Fig. \ref{casimir_energy_alpha}, we investigate the impact of the Lorentz-violating parameter, $\alpha_{\phi}$, while fixing $\beta = 0.2$. It is observed that an increase in $\alpha_{\phi}$ leads to a downward shift (making it more negative) in the Casimir energy across the entire analyzed mass spectrum. Although the parameter $\alpha_{\phi}$ is physically constrained to extremely small values by quantum gravity models \cite{Carroll:2008pk}, its asymptotic behavior for larger values reveals a divergent trend in the energy density. Now, Fig. \ref{casimir_energy_ab} illustrates the dependence of the Casimir energy on the geometric ratio $a/b$, which relates the distance between the plates to the scale of the compactified extra dimension. For small ratios ($a/b \to 0$), corresponding to the limit of large extra dimensions, the energy exhibits a sharp divergence. In contrast, for high values of $a/b$, the curves converge asymptotically to constant values, suggesting that the contribution from the compact dimension stabilizes the energy density in the large-mass regime.

\subsection{Massless Limit}

In the massless limit, $m \to 0$, the Casimir energy is obtained by applying the asymptotic expansion of the modified Bessel function $K_{\alpha}(x)$ for $x \ll 1$. This procedure yields the following expression:
\begin{align}
E_C=-\frac{L^2\pi^2}{4a^3}\left\{\frac{3}{16}\left(\frac{a}{\pi \tilde{b}}\right)^4 \sum_{n=1}^{\infty} \frac{\cos (2 \pi n \beta)}{n^5}+\frac{1}{2\pi^{4}}\left(\frac{a}{\tilde{b}}\right)^3 \sum_{n=1}^{\infty} \frac{\cos (2 \pi n \beta)}{n^4} +\frac{1}{2 \pi^2} \sum_{n=-\infty}^{\infty} \sum_{q=1}^{\infty}\left(\frac{N}{q}\right)^2 K_2(2 N q \pi)\right\}.
\label{e14}
\end{align}
Here, the first term within the brackets scales linearly with $a$. In Ref. \cite{pascoal2008estimate}, a similar study disregarded this term because it diverges in the limit $a \to \infty$. However, this term contributes as a constant to the Casimir force—the actual observable quantity. Since our objective is to investigate the influence of boundary conditions and the extra-dimension parameter on the physical force, we retain this term to analyze its experimental implications. Furthermore, for values of $a$ close to $\tilde{b}$, this linear term does not dominate the energy, ensuring a balanced contribution from all geometric factors.

Now, we consider the case $\beta \to 0$, where the standard periodic boundary condition is recovered. The result is:
\begin{equation}
\sum_{q=1}^{\infty}\left(\frac{\beta}{q}\right)^2 K_2\left(2 \beta q \pi\frac{a}{\Tilde{b}}\right) \stackrel{\beta \rightarrow 0}{\longrightarrow} \sum_{q=1}^{\infty} \frac{\Tilde{b}^2}{2a^2 \pi^2 q^4}=\frac{\pi^2\Tilde{b}^2}{180a^2} ,
\label{e15.1}
\end{equation}
and consequently, we find the following energy:
\begin{equation}
E_C=-\frac{L^2 \pi^2}{1440 a^3}\left[1+\frac{2a^3}{ \Tilde{b}^3}+\frac{135 a^4}{2 \pi^4 \Tilde{b}^4} \zeta(5)+\frac{360 a^2}{\pi^2 \Tilde{b}^2}\sum_{n=1}^{\infty} \sum_{q=1}^{\infty} \frac{n^2}{q^2} K_2\left(2 n \pi q \frac{a}{\Tilde{b}}\right)\right].
\label{e15.2}
\end{equation}
This result is in excellent agreement with Refs. \cite{pascoal2008estimate, Zhai:2014jta}. The first term in Eq. \eqref{e15.2} corresponds to the Casimir energy of a massless scalar field under Neumann boundary conditions in (3+1)D Minkowski spacetime. The second term is a correction purely due to the extra dimension; as it is independent of $a$, it does not contribute to the force between the plates. The remaining terms represent the complex interplay between the plates and the compactified geometry of the extra dimension.

\subsection{Asymptotic Analysis: Large and Small Mass Limits}

To complement the numerical results presented in the previous sections, we now examine the analytical behavior of the Casimir energy in the asymptotic limits of $M$. This analysis provides the theoretical basis for understanding the regime of light and heavy fields—corresponding to large and small plate separations, respectively—and establishes the consistent convergence to the massless case.

\subsubsection{Large Mass Regime}

In the regime where the field mass is large compared to the plate separation ($M \gg 1$), we consider the leading-order contributions to the Casimir energy. Starting from Eq. \eqref{casimir_ener_exac}, the asymptotic behavior of the modified Bessel functions $K_\nu(u) \sim \sqrt{\pi/2u} \, e^{-u}$ leads to:
\begin{equation}
E_C \approx -\frac{L^2}{16a^3}\sum_{\ell=1}^{\infty}\left\{  \cos(2\pi \ell\beta) e^{-2\pi \ell M x}\left(  \frac{ M^2}{x^{2}} \frac{1}{\ell^3}+  \frac{M^{3/2}}{x^{3/2}}\frac{1}{\ell^{5/2}}\right) + M^{3/2}\frac{e^{-2\pi \ell M}}{\ell^{5/2}} \sum_{n=-\infty}^{\infty} e^{\frac{\pi q}{M} \left( \frac{n+\beta}{x} \right)^2 }\right\} ,
\label{casimir_energy_large}
\end{equation}
where $x = \tilde{b}/a$. The first term on the right-hand side is exponentially suppressed by the factor $e^{-2\pi M x}$, while the second term, although independent of $x$ in its exponent, also decays as $e^{-2\pi M}$.

Physically, this implies two distinct exponential scales. If $x < 1$ ($\tilde{b} < a$), the factor $e^{-2\pi M x}$ decays more slowly, and the single-sum contributions dominate. Conversely, if $x > 1$ ($\tilde{b} > a$), the double-sum term becomes the dominant contribution. In both cases, the Casimir energy is exponentially suppressed for $M \gg 1$. This behavior justifies the vanishing energy observed in Figs. \ref{casimir_energy_beta}-\ref{casimir_energy_ab}.

\subsubsection{Small Mass Regime}

For this case, we consider the opposite regime where the field mass is much smaller than the geometric scales ($M \ll 1$), representing a light scalar field that is not strictly massless. Here, we perform a Taylor expansion of the modified Bessel functions for small arguments, given by:
\begin{align}
K_{5/2}(u) &\approx \frac{3 \sqrt{\pi}}{\sqrt{2}} u^{-5 / 2}-\frac{\sqrt{\pi}}{2 \sqrt{2}} u^{-1 / 2}+\mathcal{O}\left(u^{3 / 2}\right),\\
K_{2}(u) &\approx \frac{2}{u^2}- \frac{1}{2} +\frac{u^2}{16}\left(\ln\frac{u}{2} +\gamma-\frac{3}{2}\right)+\mathcal{O}(u^4 \ln u),
\end{align}
where $\gamma$ is the Euler-Mascheroni constant. Now, we can evaluate the individual contributions to the Casimir energy. Expanding each term of the energy, we find that the leading contributions are:
\begin{align}
E^1_C \approx \frac{3}{4 \pi^2} x^{-4} \sum_{n=1}^{\infty} \frac{\cos (2 \pi n \beta)}{n^5}-\frac{1}{4} M^2 x^{-2} \sum_{n=1}^{\infty} \frac{\cos (2 \pi n \beta)}{n^3}+\mathcal{O}\left(M^4 \ln M\right) , \label{cont_1} \\
E^2_C \approx \frac{1}{2\pi^2} x^{-3} \sum_{\ell=1}^{\infty} \frac{\cos(2\pi \ell \beta)}{\ell^4} -\frac{1}{2} x^{-1} M^2
\sum_{\ell=1}^{\infty} \frac{\cos(2\pi \ell \beta)}{\ell^2} + \mathcal{O}(M^4 \ln M) \label{cont_2},
\end{align}
where we consider the small-$u$ behavior of $K_{5/2}(u)$ and set $u=2\pi n M x$. Regarding the double-sum term, the leading-order expansion for small mass yields:
\begin{equation}
C^3_C \approx \sum_{n=-\infty}^{\infty} \sum_{q=1}^{\infty}\frac{N^2}{q^2}K_2(2\pi q |N| +\mathcal{O}(M^2)) \label{cont_3},
\end{equation}
where $\sqrt{M^2 + N^2} = 2\pi q |N| +\mathcal{O}(M^2)$. However, the mass-dependent terms contribute significantly when $n=0$ and $\beta \to 0$. Consequently, we obtain:
\begin{equation}
\sum_{q=1}^{\infty} \frac{M^2}{q^2} K_2(2 \pi q M)=\sum_{q=1}^{\infty}\left[\frac{1}{2 \pi^2 q^4}-\frac{M^2}{2 q^2}+\cdots\right]=\frac{\pi^2}{180}-\frac{\pi^2}{12} M^2+\cdots \ . \label{cont_4}
\end{equation}

Collecting all contributions given by Eqs. \eqref{cont_1}, \eqref{cont_2}, \eqref{cont_3} and \eqref{cont_4}, the Casimir energy for the small-mass regime is given by:
\begin{equation}
E_C \approx E_C^{\rm massless} + \frac{L^2M^2}{16a^3}\sum_{\ell=1}^{\infty}\left\{ \frac{1}{2x^2} \frac{\cos (2 \pi \ell \beta)}{\ell^3}+ \frac{1}{x} \frac{\cos(2\pi \ell \beta)}{\ell^2}+\frac{1}{\ell^2} \right\}.
\label{casimir_energy_small}
\end{equation}

This result demonstrates that in the $M \ll 1$ limit, the Casimir energy smoothly approaches the massless expression in Eq. \eqref{e14}, with mass corrections appearing at order $M^2$. This explains why the numerical curves in Figs. \ref{casimir_energy_beta}-\ref{casimir_energy_ab} converge to stable, finite values at the origin $M = 0$.

\section{Casimir force}
\label{sec4}

In this section, our focus will be the analysis of the Casimir force. Although we have obtained an analytical expression for the Casimir energy in Eq. \eqref{casimir_ener_exac}, the force acting on the plates along the $z$-direction is the physically observable quantity. From this quantity stems the possibility of testing the correction terms through experimental data. Furthermore, the force allows us to distinguish between bulk contributions and the pressure strictly induced by the boundary conditions. The Casimir force is given by:
\begin{align}
    F_C = &-\frac{\partial E_C}{\partial a}=\sum_{n=1}^{\infty}\frac{L^2 e^{-2 b m n}  \cos (2 \pi  \beta  n)}{64 \pi ^2
   b^4 n^5}[2 b m n (2 b m n+3)+3]\nonumber\\
    &-\sum_{n=-\infty}^{\infty} \sum_{q=1}^{\infty}\frac{L^2 \left(M^2+N^2\right)}{8a^4 q^2} \left[2 q \pi \sqrt{M^2+N^2} K_1\left(2 \pi  q \sqrt{M^2+N^2}\right)+3 K_2\left(2 \pi
    q \sqrt{M^2+N^2}\right)\right]\,,
    \label{e15.3}
\end{align}
where we have used the closed-form expression for the modified Bessel function of order $5/2$:
\begin{equation}
K_{5/2}(x) = \sqrt{\frac{\pi}{2x}}\, e^{-x} \left( 1+\frac{3}{x}+\frac{3}{x^{2}} \right).
\end{equation}
Note that in Eq. \eqref{e15.3}, the first term represents the pure contribution from the compactified extra dimension, as it is independent of the distance between the plates, $a$.

\begin{figure}[h!]
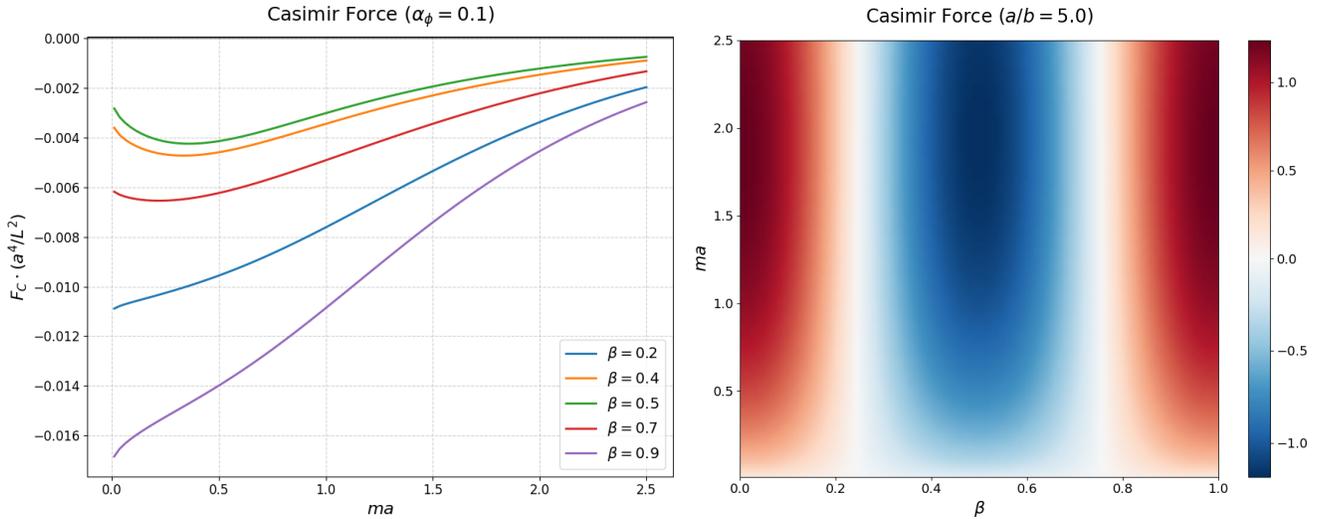

    \includegraphics[scale=0.40]{force_casimir_beta.pdf}
    \includegraphics[width=0.465\linewidth]{force_casimir_beta_hm.pdf}
    \caption{Casimir force as a function of $ma$ for different values of the quasi-periodicity parameter $\beta$. The $\beta$ acts as a control parameter, capable of modulating the force's magnitude and sign, resulting in attractive, repulsive, or null regimes.}
    \label{casimir_force_beta}
\end{figure}

\begin{figure}[h!]
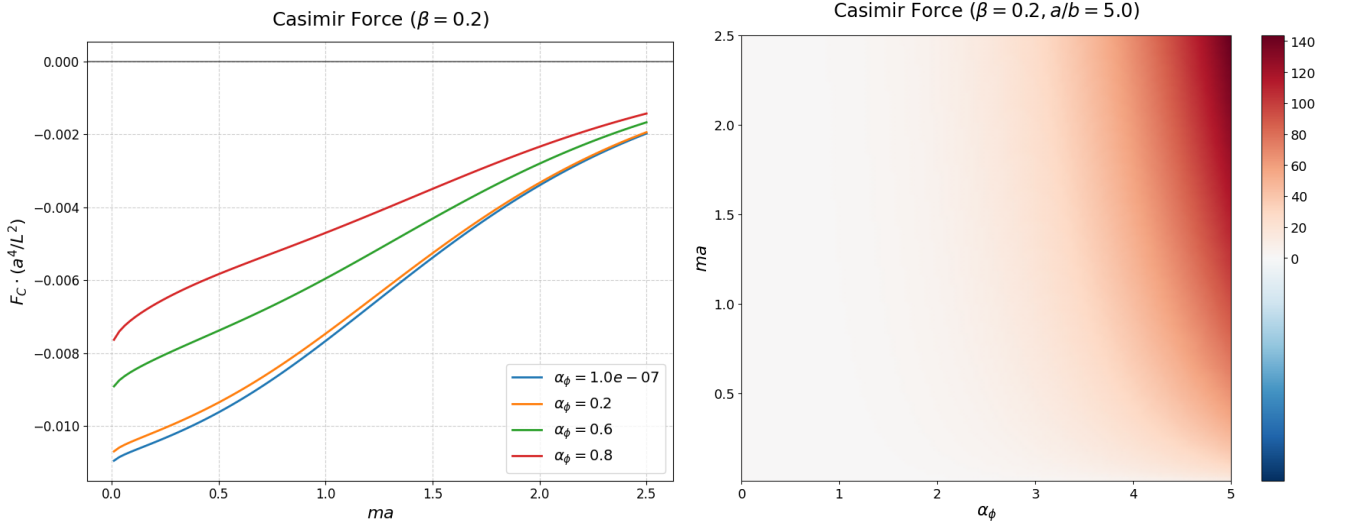

    \includegraphics[scale=0.40]{casimir_force_alpha.pdf}
    \includegraphics[width=0.47\linewidth]{casimir_force_alpha_hm.pdf}
    \caption{Casimir force as a function of $ma$ for several values of the LV parameter $\alpha_\phi$. The plot reveals that increasing $\alpha_\phi$ enhances the force magnitude across the mass spectrum and divergent behavior for large violation scales.}
    \label{casimir_force_alpha}
\end{figure}

\begin{figure}[h!]
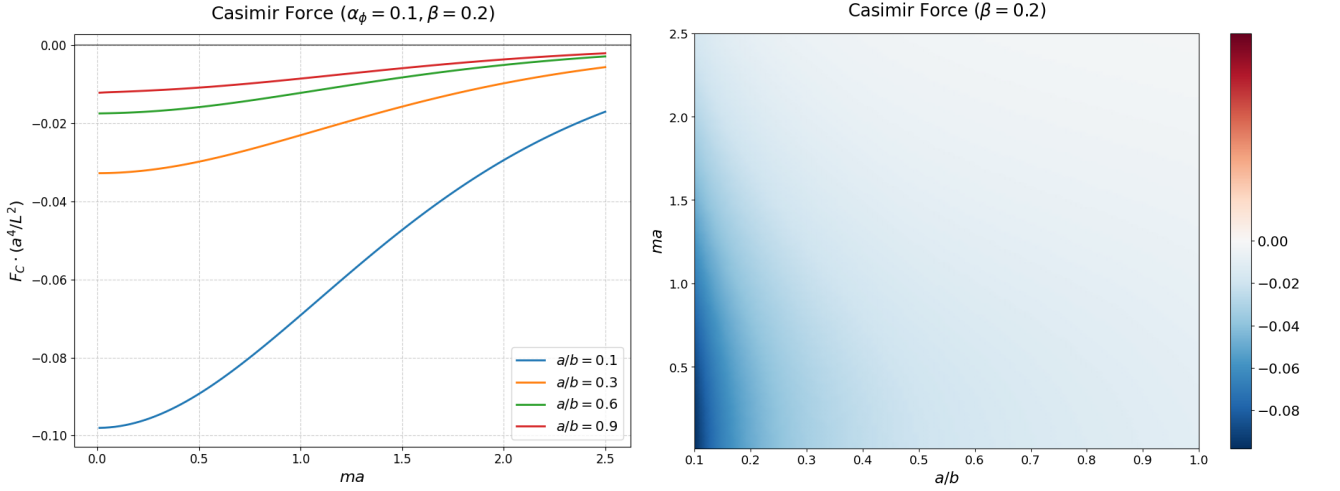

    \includegraphics[scale=0.375]{casimir_force_ab.pdf}
    \includegraphics[width=0.49\linewidth]{casimir_force_ab_hm.pdf}
    \caption{Casimir force as a function of $ma$ for different geometric ratios $a/b$. The plot highlights the transition from a strongly attractive regime at small ratios to a dominant repulsive regime in the high-compactification limit.}
    \label{casimir_force_ab}
\end{figure}

Following the same procedure applied to the Casimir energy, we numerically investigate the Casimir force, as defined in Eq. \eqref{e16}. The plots illustrate the effects of varying the parameters $\beta$, $\alpha_{\phi}$, and the geometric ratio $a/b$. As evidenced by Fig. \ref{casimir_force_beta}, the quasi-periodic parameter $\beta$ modulates both the intensity and the nature of the force—allowing for attractive, repulsive, or null configurations. This behavior is in agreement with Refs. \cite{deFarias:2020xms, deFarias:2021qdg}, further supporting the role of $\beta$ as a crucial control parameter. Furthermore, Fig. \ref{casimir_force_alpha} shows that the force magnitude increases significantly as the LV parameter $\alpha_{\phi}$ grows, indicating a formal divergence in the limit of large Lorentz violation. Regarding the geometric influence, Fig. \ref{casimir_force_ab} illustrates the Casimir force in terms of the ratio $a/b$. It is evident that for small values of $a/b$, the force diverges toward negative infinity, indicating a strong attractive regime. Conversely, for large values of this ratio, the force asymptotically approaches a constant positive value, suggesting that the repulsive contribution becomes dominant in the high-compactification limit.

\subsection{Massless Limit}

Next, by taking the massless limit ($m \to 0$), the Casimir force expression in Eq. \eqref{e15.3} simplifies to:
\begin{align}
 F_C=&-\frac{L^2}{128 \tilde{b}^4} \sum_{\delta=+,-}\left\{-\frac{3}{\pi^2} Li_5\left(e^{2 i \pi\beta\delta}\right)\right.\nonumber \\
& +16\sum_{n=1}^{\infty} \sum_{q=1}^{\infty}\left[3\left(\frac{ \tilde{b}(\delta n+\beta)}{aq}\right)^2 K_2\left(\frac{a}{\tilde{b}} 2 \pi q|\delta n+\beta|\right)+2\pi\left.\frac{ \tilde{b} (\delta n+\beta)^3}{a q} K_1\left(\frac{a}{\tilde{b}} 2 \pi q|\delta n+\beta|\right)\right]\right\}\nonumber \\
&-\frac{L^2}{8 \tilde{b}^4}\sum_{q=1}^{\infty}\left[3\left(\frac{ \tilde{b}\beta}{aq}\right)^2 K_2\left(\frac{a}{\tilde{b}} 2 \pi q\beta\right)+2\pi\frac{ \tilde{b} \beta^3}{a q} K_1\left(\frac{a}{\tilde{b}} 2 \pi q\beta\right)\right].
\label{e16}
\end{align}
Now, we examine two fundamental cases: the periodic ($\beta=0$) and antiperiodic ($\beta=1/2$) conditions. For the periodic case, the Casimir force can be expressed as:
\begin{equation}
    F_C=-\frac{L^2 \pi^2}{480a^4}+\frac{3L^2 }{64 \Tilde{b}^4\pi^2} \zeta(5)- \frac{L^2\pi}{2a\Tilde{b}^3}\sum_{n=1}^{\infty} \sum_{q=1}^{\infty}\frac{ n^3}{q} K_1\left(2\pi n q \frac{a}{\Tilde{b}}\right)- \frac{3L^2}{(2a\Tilde{b})^2}\sum_{n=1}^{\infty} \sum_{q=1}^{\infty}\left(\frac{ n}{q} \right)^2 K_2\left(2\pi n q \frac{a}{\Tilde{b}}\right),
   \label{e20}
\end{equation}
where we have made use of Eq. \eqref{e15.1}. The above result agrees with the expression obtained in Ref. \cite{pascoal2008estimate}. Note that, by taking the limit $a \to \infty$, we obtain the contribution of the extra dimension to the Casimir force, given by:
\begin{equation}
    F_C=\frac{12\zeta(5)L^2 }{\pi^2b^4}(1+\alpha_{\phi})^2 .
\end{equation}
However, as $a \to 0$, the Casimir force in Eq. \eqref{e20} exhibits the expected divergence, consistent with the short-distance behavior of vacuum fluctuations.

For the antiperiodic case, we obtain:
\begin{align}
F_C= & -\frac{45 L^2}{1024 \Tilde{b}^4} \zeta(5)-\frac{L^2}{32 a \Tilde{b}^3} \sum_{\delta=+,-}\sum_{n=1}^{\infty} \sum_{q=1}^{\infty}\left\{\frac{\pi}{q}(2 \delta n+1)^3 K_1\left[\frac{a}{\Tilde{b}} \pi q|2 \delta n+1|\right]\right.\nonumber \\
& \left.+\frac{\Tilde{b}(2 \delta n+1)^2}{3 a q^2} K_2\left[x \pi q(2 \delta n+1)\right]\right\} -\frac{L^2}{32 a \Tilde{b}^3} \sum_{q=1}^{\infty}\left[\frac{\pi}{q} K_1\left(\frac{a \pi q}{\Tilde{b}}\right)+\frac{3 \Tilde{b}}{a q^2} K_2\left(\frac{a \pi q}{\Tilde{b}}\right)\right].
\label{e21}
 \end{align}
In this case, the force diverges negatively as $a \to 0$ and remains attractive for all values of $a$. Hence, the presence of the extra-dimension parameter $\tilde{b}$ significantly modifies the interaction, maintaining an attractive profile between the plates.

\subsection{Asymptotic Analysis: Large and Small Mass Limits}

Analogously to the analysis performed for the Casimir energy, we now examine the analytical behavior of the Casimir force in the asymptotic regimes.

\subsubsection{Large Mass Limit}

In the limit of large field mass ($M \gg 1$), the force between the plates is governed by the asymptotic decay of the vacuum fluctuations. By evaluating the derivative of the Casimir energy in this regime or directly expanding Eq. \eqref{e15.3} using the asymptotic form of the Bessel functions, we obtain:
\begin{equation}
F_C \approx -\frac{L^2 \pi M^3}{8 a^3 \tilde{b}} \sum_{\ell=1}^{\infty} \frac{\cos(2\pi \ell \beta)}{\ell^2} e^{-2\pi \ell M \frac{\tilde{b}}{a}}.
\label{casimir_force_large}
\end{equation}
Physically, this indicates that as the field becomes sufficiently massive, the vacuum fluctuations are unable to mediate a significant long-range interaction between the plates, leading to a decoupling effect. This analytical result is consistent with the behavior observed in Figs. \ref{casimir_force_beta}, \ref{casimir_force_alpha}, and \ref{casimir_force_ab}, where the force asymptotically vanishes for large values of $M$.

\subsubsection{Small Mass Limit}

For the regime of light fields, we expand the Casimir force in powers of the dimensionless mass parameter, $M \ll 1$. Utilizing the small-argument expansions of the modified Bessel functions $K_1(u)$ and $K_2(u)$, the leading-order corrections to the Casimir force can be expressed as:
\begin{equation}
F_C \approx F_C^{\rm massless} + \frac{L^2 M^2}{16 a^4} \sum_{\ell=1}^{\infty} \left\{ \frac{1}{2x^2} \frac{\cos(2\pi \ell \beta)}{\ell^3} + \frac{2}{x} \frac{\cos(2\pi \ell \beta)}{\ell^2} + \frac{3}{\ell^2} \right\}.
\label{force_casimir_small}
\end{equation}
In this limit, the force smoothly approaches the massless expression given in Eq. \eqref{e16}. Therefore, the presence of the mass introduces a correction of order $(ma)^2$, which shifts the force magnitude.

So, this analytical behavior confirms the stability of the numerical solutions at the near origin ($M=0$) and ensures that for light particles, such as those predicted in certain dark matter or axion-like models, the Casimir force retains a profile nearly identical to the massless case, with small deviations proportional to the square of the mass.
\\
\section{Conclusion}
\label{sec5}

In this work, we investigated the influence of Lorentz symmetry breaking, introduced by the aether-like field $\alpha_\phi$. Furthermore, we considered the scenario of a fifth dimension in flat spacetime, in conjunction with a quasiperiodic condition given by the parameter $\beta$. The presence of this extra dimension, compactified at a scale $b$, modifies the standard dispersion relation and, together with the Neumann boundary conditions applied to parallel plates, gives rise to a modified Casimir effect.

We derived a closed-form expression for the Casimir energy, Eq. \eqref{casimir_ener_exac}, explicitly showing its dependence on the plate separation $a$, the compactification scale $b$, the Lorentz-violating parameter $\alpha_{\phi}$, and the quasiperiodicity $\beta$. Our numerical analysis in Fig. \ref{casimir_energy_beta} revealed that $\beta$ acts as a control parameter, allowing the energy to transition between attractive and repulsive regimes. Specifically, we observed a symmetric behavior around $\beta = 0.5$, whereas the energy remains strictly negative in the periodic limit ($\beta \to 0$).

Furthermore, Fig. \ref{casimir_energy_alpha} demonstrates that the aether-like field $\alpha_\phi$ acts as an enhancement factor for the Casimir energy, where an increase in the LV parameter results in a downward shift of the energy levels. Although such violation is phenomenologically restricted to minute scales, our results indicate that, in the asymptotic limit, $\alpha_\phi$ triggers a divergent trend that significantly amplifies the vacuum energy. As illustrated in Fig. \ref{casimir_energy_ab}, the geometric configuration and the ratio $a/b$ prove to be decisive factors in the stabilization of the system. While the limit of large extra dimensions ($a/b \to 0$) leads to a sharp divergence, the high-compactification regime ensures that the contribution from the fifth dimension plateaus, effectively stabilizing the vacuum interaction across the mass spectrum.

Regarding the Casimir force, we obtained a closed-form expression that characterizes the interaction between the plates, as given in Eq. \eqref{e15.3}. Similarly to the Casimir energy, our results of the Figs. \ref{casimir_force_beta}, \ref{casimir_force_alpha} and \ref{casimir_force_ab} demonstrate that the force's intensity and sign are highly sensitive to the geometric ratio $a/b$ and the parameters $\beta$ and $\alpha_{\phi}$. In the high-compactification limit ($a/b \to \infty$), the force asymptotically approaches a constant value, representing a stabilization of the vacuum pressure. Furthermore, the analysis of periodic ($\beta = 0$) and antiperiodic ($\beta = 1/2$) cases confirms the consistency of our model with established theoretical benchmarks.

For the massless limit ($m \to 0$), our results in Eqs. \eqref{e14} and \eqref{e16} show that the Casimir energy and force smoothly transition to their standard geometric forms. For light fields ($M \ll 1$), the leading-order mass corrections, as expressed in Eqs. \eqref{casimir_energy_small} and \eqref{force_casimir_small}, demonstrate that the interaction retains its fundamental profile, exhibiting only subtle quadratic deviations. Conversely, in the heavy-field regime ($M \gg 1$), the analytical behavior derived in Eqs. \eqref{casimir_energy_large} and \eqref{casimir_force_large} reveals a rapid exponential suppression of the Casimir effect.

As a natural extension of this work, the inclusion of finite temperature effects could reveal a complex phase diagram for the Casimir interaction, where thermal corrections might compete with Lorentz-violating contributions at different length scales.

{\acknowledgments}

The authors would like to thank CNPq and CAPES for partial financial support. K.E.L.F. is grateful to CNPq for financial support. M.A.A. and E.P. acknowledge support from CNPq (Grant nos. 306398/2021-4, 304852/2017-1, and Universal 406875/2023-5). Additionally, A.A. Araújo Filho is supported by CNPq and FAPESQ (Project nos. 150223/2025-0 and 1951/2025).

\appendix
\section{Summation Identity}
\label{app:sum_identity}

In this appendix, we present the summation identity used to evaluate the Casimir energy in the previous sections. Following the methodology discussed in Ref. \cite{deFarias:2023xjf}, the sum over the quasi-periodic modes is expressed as:
\begin{align}
\sum_{n=-\infty}^{\infty}\left[(n+\beta)^2+\mu^2\right]^{-s} &= 2 \int_0^{\infty} \mathrm{d} x\left[x^2+\mu^2\right]^{-s} \nonumber \\
&+ \left(i^{1-2s}+(-i)^{1-2s}\right) \sum_{\ell=1}^{\infty} 2 \cos (2 \pi \ell \beta) \int_\mu^{\infty} d u e^{-2 \pi \ell u}\left[u^2-\mu^2\right]^{-s} \nonumber \\
&= \sqrt{\pi} \mu^{1-2 s} \frac{\Gamma\left(s-\frac{1}{2}\right)}{\Gamma(s)} + \left(i^{1-2s}+(-i)^{1-2s}\right) \frac{2\Gamma(1-s)}{\pi^{1-s}} \sum_{\ell=1}^{\infty}\left(\frac{\ell}{\mu}\right)^{s-\frac{1}{2}} \cos (2 \pi \ell \beta) K_{\frac{1}{2}-s}(2 \pi \mu \ell) \nonumber \\
&= \sqrt{\pi} \mu^{1-2 s} \frac{\Gamma\left(s-\frac{1}{2}\right)}{\Gamma(s)} + \frac{4\Gamma(1-s)\sin(\pi s)}{\pi^{1-s}} \sum_{\ell=1}^{\infty}\left(\frac{\ell}{\mu}\right)^{s-\frac{1}{2}} \cos (2 \pi \ell \beta) K_{\frac{1}{2}-s}(2 \pi \mu \ell).
\label{appendix_1}
\end{align}

This identity allows for a clear separation between the vacuum contribution of the free space (first term) and the finite, $\ell$-dependent terms arising from the compactification and quasi-periodicity (second term). In the context of the Casimir effect, the first term is typically removed via renormalization, leaving only the finite summation involving the modified Bessel functions of the second kind, $K_{\nu}$.


\end{document}